\documentclass{PoS}

\title{Results and Prospects from NOvA}

\ShortTitle{Results and Prospects from NOvA}

\author{\speaker{Jianming Bian (for the NOvA Collaboration)}\thanks{}\\
        University of California, Irvine\\
        E-mail: \email{bianjm@uci.edu}}

\abstract{NOvA is a long-baseline neutrino experiment that uses an upgraded NuMI neutrino source at Fermilab and a 14-kton detector at Ash River, Minnesota. The detector has a highly active,
finely segmented design that offers superb event identification capability. This talk presents the latest  $\nu_\mu$ ($\bar{\nu}_\mu$) disappearance and $\nu_e$ ($\bar{\nu}_e$) appearance combined results using the first NOvA anti-neutrino beam data. In the far detector, 18 $\bar{\nu}_e$ candidate events are observed, with a significance of $\bar{\nu}_e$ appearance more than 4 $\sigma$. The NOvA results favor a normal neutrino mass hierarchy.}

\FullConference{The 20th International Workshop on Neutrinos from Accelerators (NuFACT2018) \\
		12-18 August 2018\\
		Blacksburg, Virginia}

\begin{document}

\section{Introduction}

NOvA is a major U.S.-based long-baseline neutrino experiment  optimized to observe the oscillation of muon neutrinos to electron-neutrinos.  It uses a 14-kt liquid scintillator Far Detector (FD) in Ash River, Minnesota, to detect the oscillated NuMI (Neutrinos at the Main Injector) muon neutrino beam produced at Fermilab. The NuMI beam has been upgraded to 700 kW.    The NOvA baseline is 810 km, the longest in operation, which enhances the matter effect and allows probing of the neutrino mass ordering. NOvA is equipped with a 0.3-kt functionally identical Near Detector (ND) located at Fermilab to measure un-oscillated beam neutrinos and estimate backgrounds at the FD.  Both the FD and ND are located 14.6 mrad off-axis to receive a narrow-band neutrino energy spectrum near 2 GeV, around the $\nu_\mu\to\nu_e$ oscillation maximum. The off-axis design enhances the $\nu_\mu\to\nu_e$ oscillation signal in the FD while reducing beam backgrounds from unoscillated neutrinos. 

The NOvA detectors are fine-grained and highly active tracking calorimeters. They consist of plastic (PVC) extrusions filled with liquid-scintillator, with wavelength-shifting fibers (WLS) connected to avalanche photodiodes (APDs).  The detector cells' cross-sectional size is about 6 cm $\times$ 4 cm. Each cell extends the detector's full width or height, 15.6 meters in the far and 4.1 meters in the near detector. Cells are assembled in alternating layers of vertical and horizontal extrusions, so cell hits caused by neutrino events are recorded in an x-z view and a y-z view hit map,  and the detectors can both measure calorimetric energy and reconstruct 3-D tracks. The NOvA detectors have low-Z and low-density, each plane is just 0.15 $X_0$, which is great for $e/\pi^0$ separation. 

\begin{figure}[h]
    \centering
    \includegraphics[width=12cm]{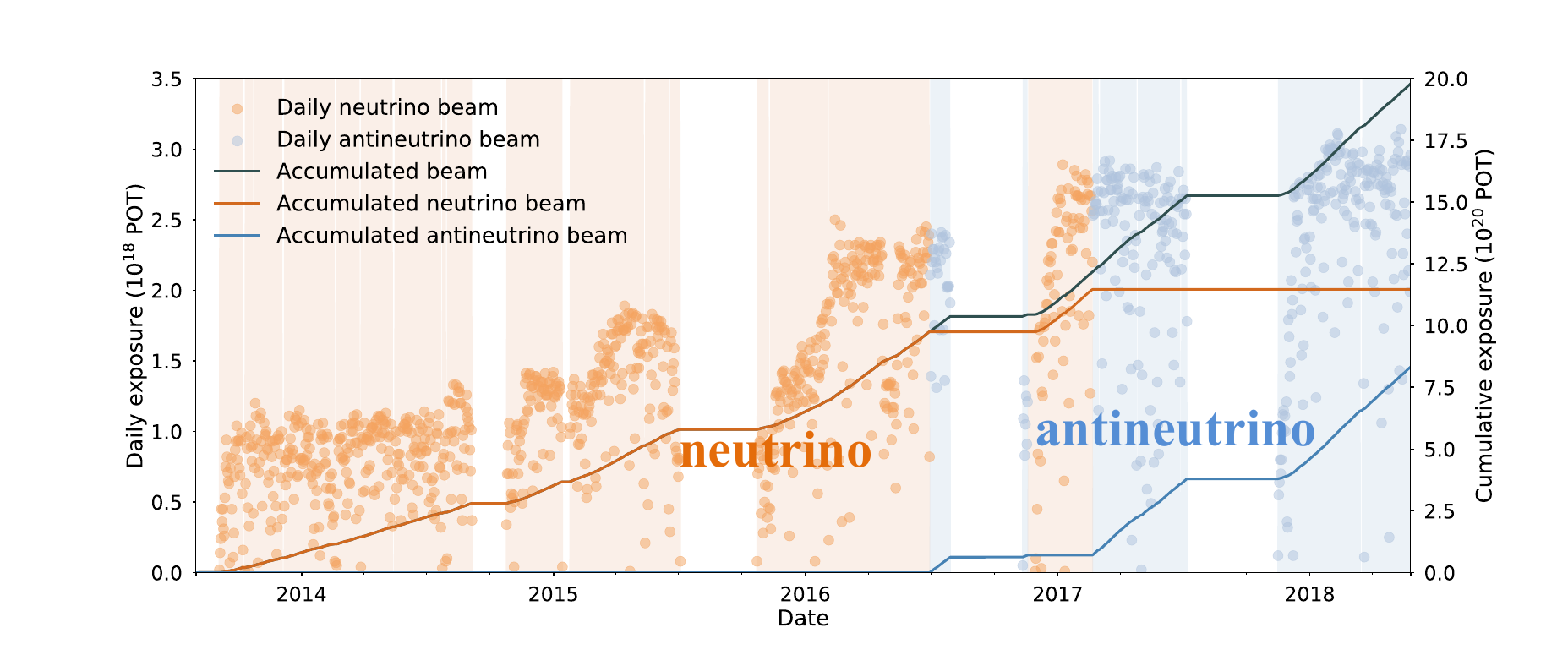}
    \caption{Time series showing the daily neutrino beam (orange) or antineutrino beam (blue) POT recorded by NOvA, from the start of commissioning to 2018-05-27. Also plotted are lines for the cumulative neutrino beam POT (dark orange), cumulative antineutrino beam POT (dark blue) and total accumulated POT (grey).}
    \label{fig:novabeam}    
    \end{figure}

\begin{figure}[h]
    \centering
    \includegraphics[width=6cm]{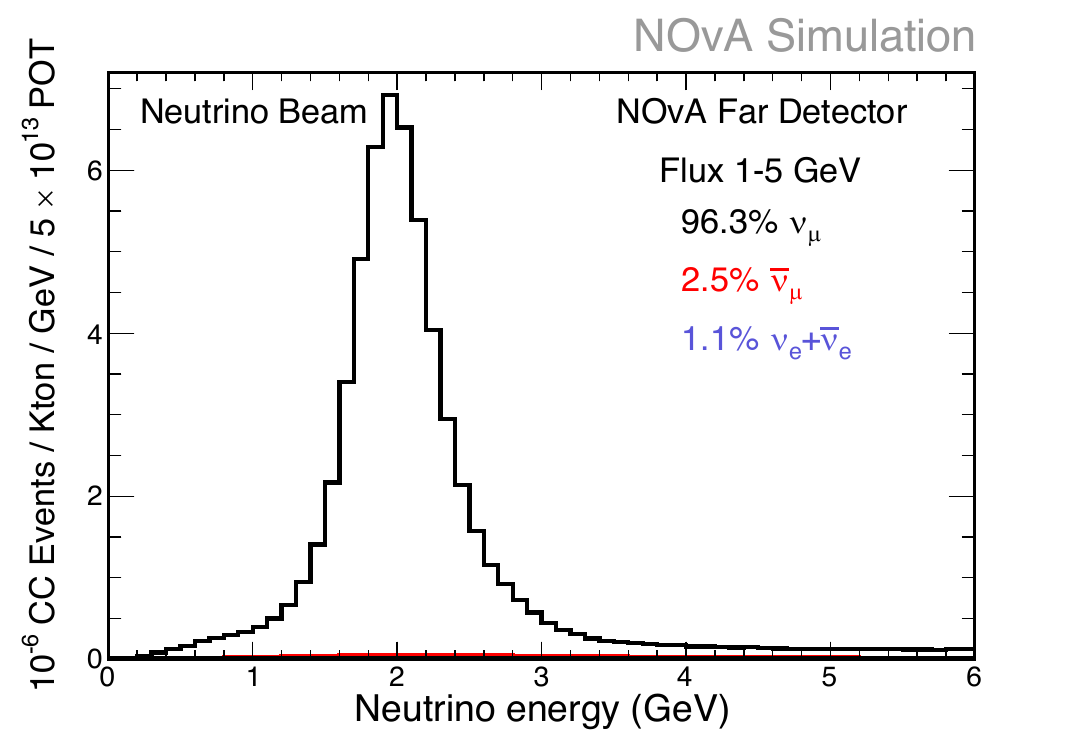}
     \includegraphics[width=6cm]{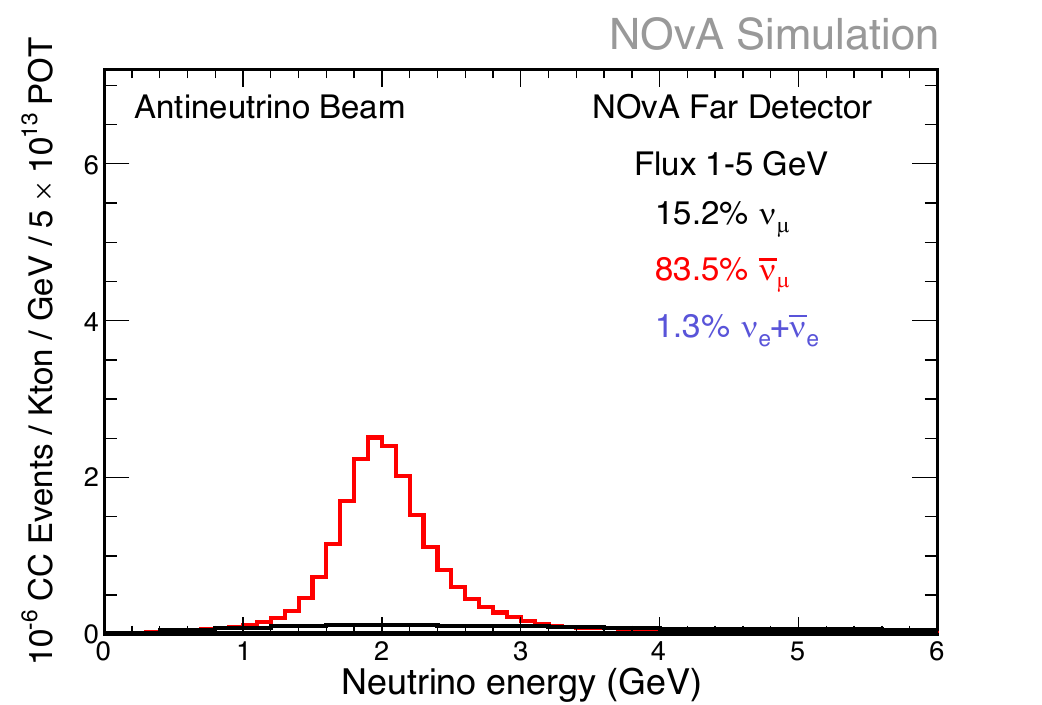}
    \caption{Charged Current (CC) event rates at FD in (left) neutrino beam and (right) anti-neutrino beam.}
    \label{fig:nubeam}
    \end{figure}

The (anti-)$\nu_\mu$ disappearance and (anti-)$\nu_e$ appearance combined analysis at NOvA aims to determine the neutrino mass hierarchy, CP violation and the octant of $\theta_{23}$. For this analysis, we use the first anti-neutrino beam data (RHC) taken from Feb 2017 to April 2018 ($6.9\times10^{20}$ POT) and the previous neutrino beam data (FHC) taken from Feb 2014 to Feb 2017 ($8.85\times10^{20}$ POT)~\cite{Adamson:2017gxd} to measure the probabilities of (anti-)$\nu_\mu\to\nu_\mu$ and (anti-)$\nu_\mu\to\nu_e$. This is the first time we include information from $\bar{\nu}_\mu\to\bar{\nu}_e$ and  $\bar{\nu}_\mu\to\bar{\nu}_\mu$ in NOvA's oscillation measurements. Figure~\ref{fig:novabeam} shows daily exposures of neutrino beam and antineutrino beam data recorded by NOvA. Figure~\ref{fig:nubeam}  shows Charged Current (CC) event rates at NOvA's FD in neutrino (left) and anti-neutrino (right) beams. 

\section{Simulation and Reconstruction\label{sec:simrec}}

\begin{figure}[h]
    \centering
    \includegraphics[width=7cm]{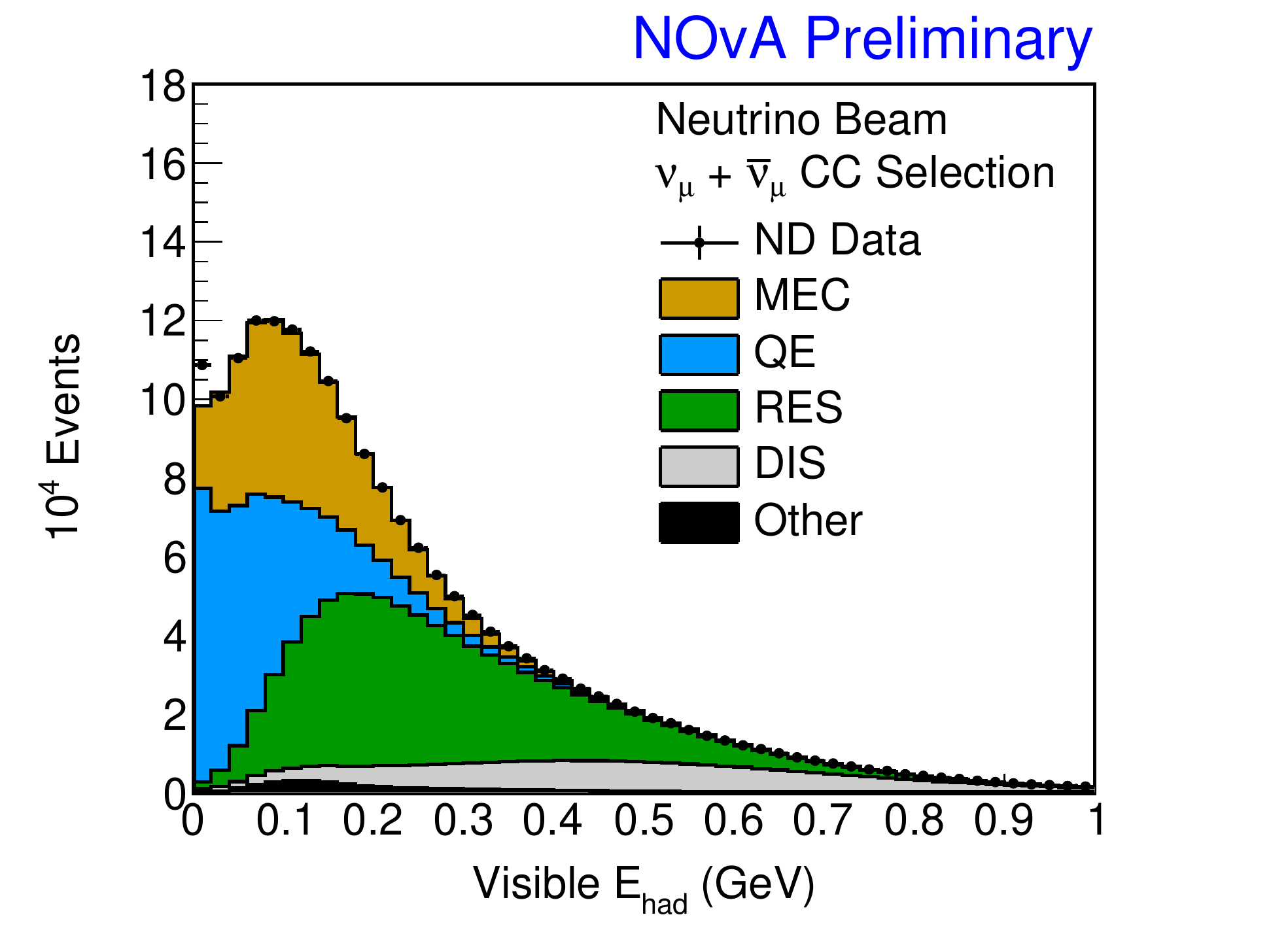}
     \includegraphics[width=7cm]{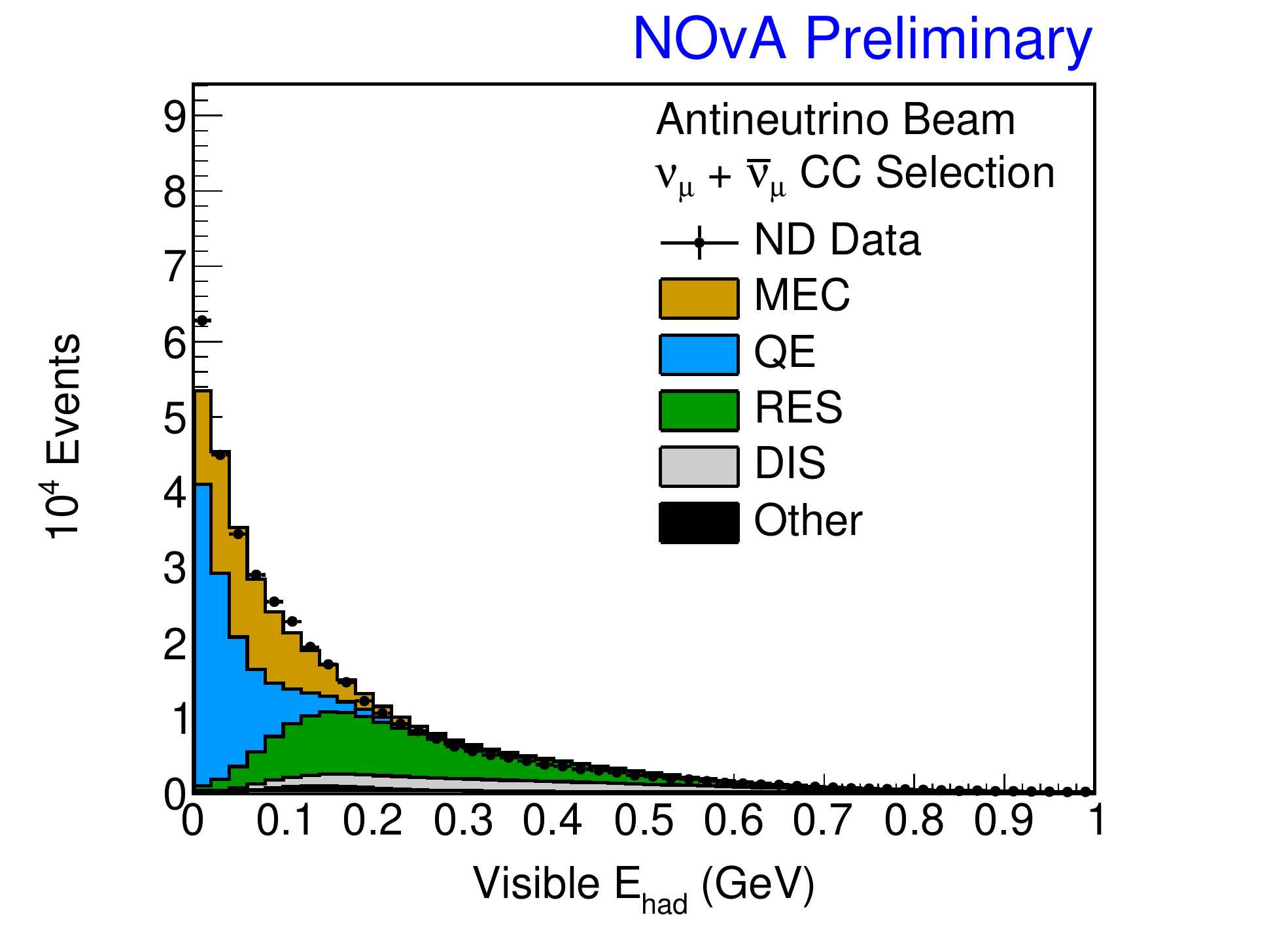}
    \caption{The near detector reconstructed visible hadronic energy distributions with GENIE cross section tuning for (left) neutrino beam data and (right) antineutrino beam data.}
    \label{fig:ndxsec}
    \end{figure}

At NOvA, the beam is simulated by GEANT4~\cite{ref:GEANT4} and corrected according to external thin-target hadroproduction data with the PPFX tool~\cite{Aliaga:2016oaz}. The simulation of NuMI neutrino beam is described in Ref.~\cite{NOvA:2018gge}.  Interactions of neutrinos on nucleus in NOvA detectors are simulated by GENIE~\cite{ref:GENIE} with several tunings to improve the Data/MC consistencies: (1) QE, RES interactions are tuned to consider long-range nuclear correlations using València model via work of R. Gran (MINERvA)~\cite{Gran:2017psn}; (2) DIS events at high invariant mass (W$>$1.7 GeV/$c^2$) are weighted up $10\%$ based on NOvA data; (3) Empirical MEC (Meson Exchange Current) model is used for Multi-nucleon ejection (2p2h)~\cite{ref:2p2h}, with the amount tuned in 2D 3-momentum and energy transfers space ($q_0 =E_\nu-E_\mu$ and $|q|=|p_\nu-p_\mu|$)  to match ND data. Systematic uncertainties in the shape of MEC events are estimated by re-fitting using alternative models with QE and RES related systematic shifts. Figure~\ref{fig:ndxsec} shows reconstructed visible hadronic energy distributions in ND after the GENIE cross section tuning. The NOvA detector responses are simulated by GEANT4. The customized NOvA detector simulation chain is described in~\cite{Aurisano:2015oxj}.

The reconstruction chain at NOvA is described in \cite{ref:reco0}. In each 550-$\mu$s trigger window, different neutrino interactions in the two detector views are separated by clustering~\cite{ref:reco1}. In this step, cell hits are clustered by space and time. This clustering process separates neutrino interactions caused by beam neutrinos from cosmic ray interactions in a time window. The procedure collects cell hits from a single neutrino interaction (slice). The slices then serve as the foundation for all later reconstruction stages.  In order to reconstruct the vertex position for each slice, a modified Hough transform is used to fit straight-lines to cell hits. Then the lines are tuned in an iterative procedure until they converge to the vertex. Electromagnetic (EM) and hadronic showers are reconstructed based on distances from hits to the lines associated with each of the particles that paths emanating from the reconstructed vertex. EM and hadronic showers are identified with a convolutional neural network. Muon tracks are fitted by a Kalman filter and recognized based on $dE/dx$, track length and scattering~\cite{ref:reco2}-\cite{ref:mureco}.

A convolutional neural network based algorithm (CVN) has been implemented at NOvA to serve as the primary event identifier~\cite{Aurisano:2016jvx}.  It uses pixels as inputs and the output is a variable that describes the probability to be $\nu_e$ ($\bar{\nu}_e$) CC, $\nu_\mu$ ($\bar{\nu}_\mu$) CC, NC and cosmic ray  event with a range $0-1$. This neural network uses convolutional filters to automatically extract features from the raw hit map inputs in the two detector views. CVN is trained with neutrino and antineutrino beam MC separately, mixed with cosmic ray data. The statistical power of the convolutional neural network method is equivalent to $30\%$ more exposure than previous PIDs, LID~\cite{Bian:2015opa} and LEM~\cite{Backhouse:2015xva}.  At NOvA, CVN has been extended to single particle identification (prong CVN)~\cite{Psihas:2018czu}. Recently, a regression convolutional neural network was developed to reconstruct electron and $\nu_e$ CC energy for future NOvA analyses~\cite{Baldi:2018qhe}.

\section{Event Selection and Neutrino Energy}

For both $\nu_\mu$  and $\nu_e$ analyses, a series of data quality requirements are used to select data taken under the normal beam and detector conditions. Events in a 12-$\mu$s timing window around the neutrino beam spill time are selected to reject cosmic rays collected in the 550-$\mu$s trigger window. Prior to the CVN selections, reconstruction quality and containment cuts are applied to remove particles that are poorly reconstructed or cosmic rays (rock muons) that enter from the edges of the FD (ND). The CVN identifier is used to select $\nu_\mu$ ($\bar{\nu}_\mu$) CC and $\nu_e$ ($\bar{\nu}_e$) CC events in the ND and FD. Cosmic ray events in the FD $\nu_\mu$ ($\bar{\nu}_\mu$) sample are rejected by a boosted decision tree (BDT) algorithm using PID and location information. The $\nu_e$ ($\bar{\nu}_e$) sample in the FD includes events in a peripheral region with less stringent containment selection to improve statistics. A location dependent BDT and tighter CVN cuts are applied to select  $\nu_e$ ($\bar{\nu}_e$)  signal events in this peripheral sample.  In the FD, the overall $\nu_\mu$ ($\bar{\nu}_\mu$) CC selection efficiency is $31.2\% (33.9\%)$ and the $\nu_e$ ($\bar{\nu}_e$) CC efficiency is $62\% (67\%)$ for the neutrino (anti-neutrino) beam. Signal purities in FD are $98.6\% (98.8\%)$ for $\nu_\mu$ ($\bar{\nu}_\mu$) and $62\% (67\%)$ for $\nu_e$ ($\bar{\nu}_e$). Wrong sign contamination in the selected  $\nu_\mu$ ($\bar{\nu}_\mu$) events in ND is $3\% (11\%)$ for neutrino (antineutrino) beam. 

\begin{figure}[h]
    \centering
    \includegraphics[width=8cm]{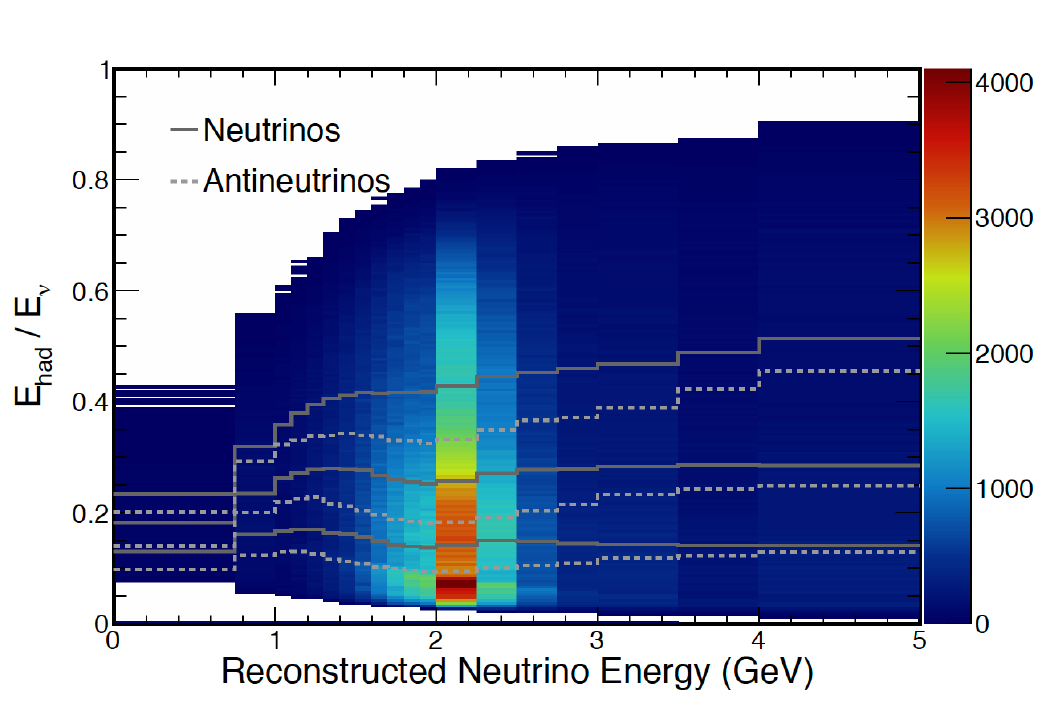}
    \caption{The four $\nu_\mu$ energy quantiles based on $E_{had}/E_\nu$.}
    \label{fig:numuquartiles}
    \end{figure}

The $\nu_\mu$ energy is reconstructed as the sum of muon energy and the hadronic energy. The muon energy is estimated from the track length and the hadronic energy is estimated via the sum of calorimetric energies excluding the muon track energy. To improve the $\nu_\mu$ energy resolution, the selected  $\nu_\mu$ sample is divided into four equal energy quantiles based on $E_{had}/E_{\nu}$, as shown in Figure~\ref{fig:numuquartiles}. The overall $\nu_\mu$ energy resolution is $8.1\% (9.1\%)$ in the FD for the neutrino (antineutrino) beam. The $\nu_\mu$ ($\bar{\nu}_\mu$) energy resolution in each quantiles varies from $5.8\% (5.5\%)$ to $11.7\% (10.8\%)$.

%resolution varies from 5.8% (5.5%) to 11.7% (10.8%) for neutrino (antineutrino) beam.

The $\nu_e$ energy estimator is based on a quadratic function of the reconstructed electron and hadronic energy ~\cite{ref:taenergy}. Parameters of the quadratic function are determined using simulated data. The electromagnetic energy component is estimated by the sum of calorimetric energies from the electron shower. The hadronic energy is estimated via the sum of calorimetric energies excluding the electron shower energy. Electrons, photons, and hadrons are identified by the prong CVN.  The  $\nu_e$ ($\bar{\nu}_e$) energy resolution is  $8.8\% (10.7\%)$ in FD for the neutrino (antineutrino) beam.

\section{Data Analysis and Systematic Uncertainties}

\begin{figure}[h]
    \centering
    \includegraphics[width=6cm]{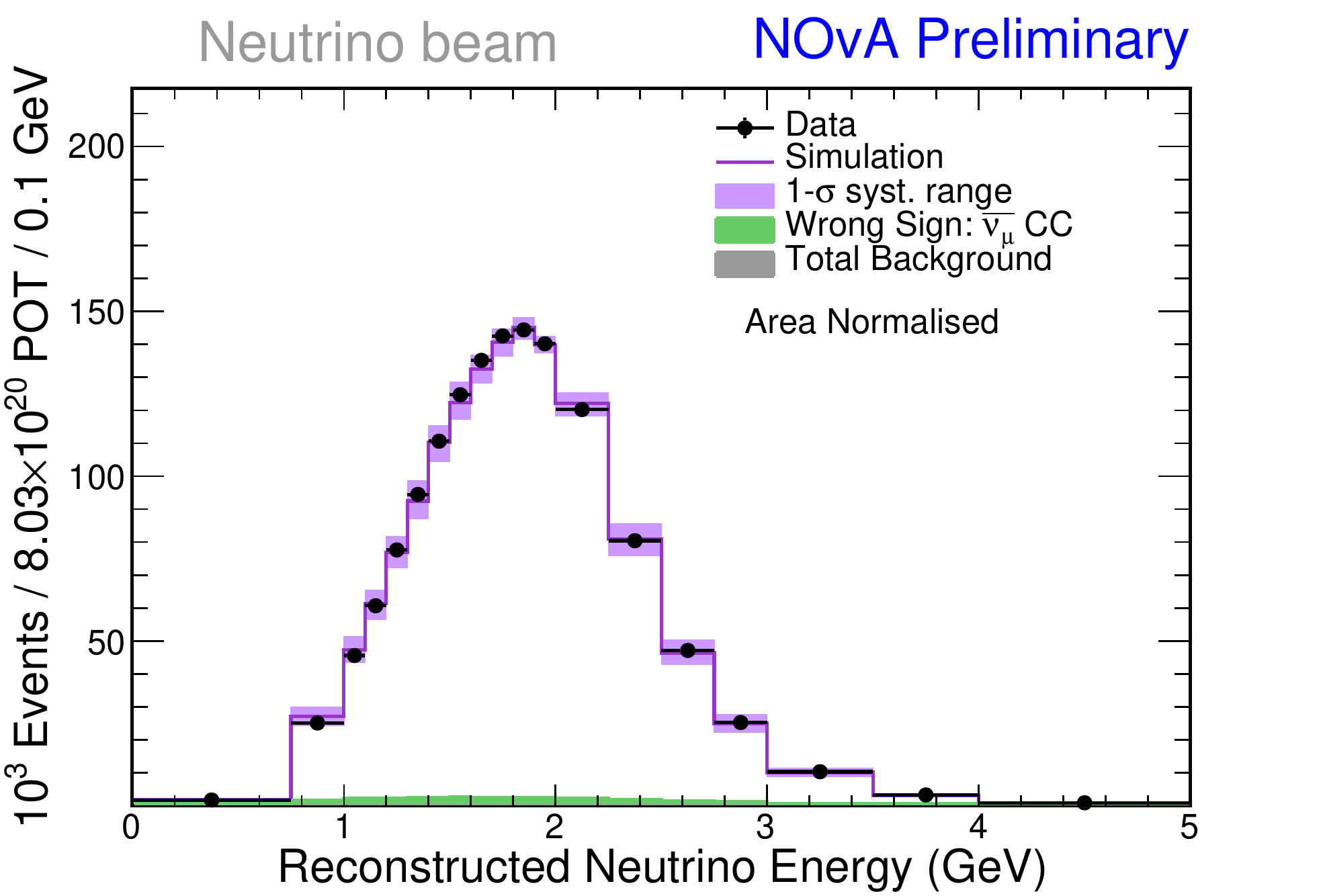}
     \includegraphics[width=6cm]{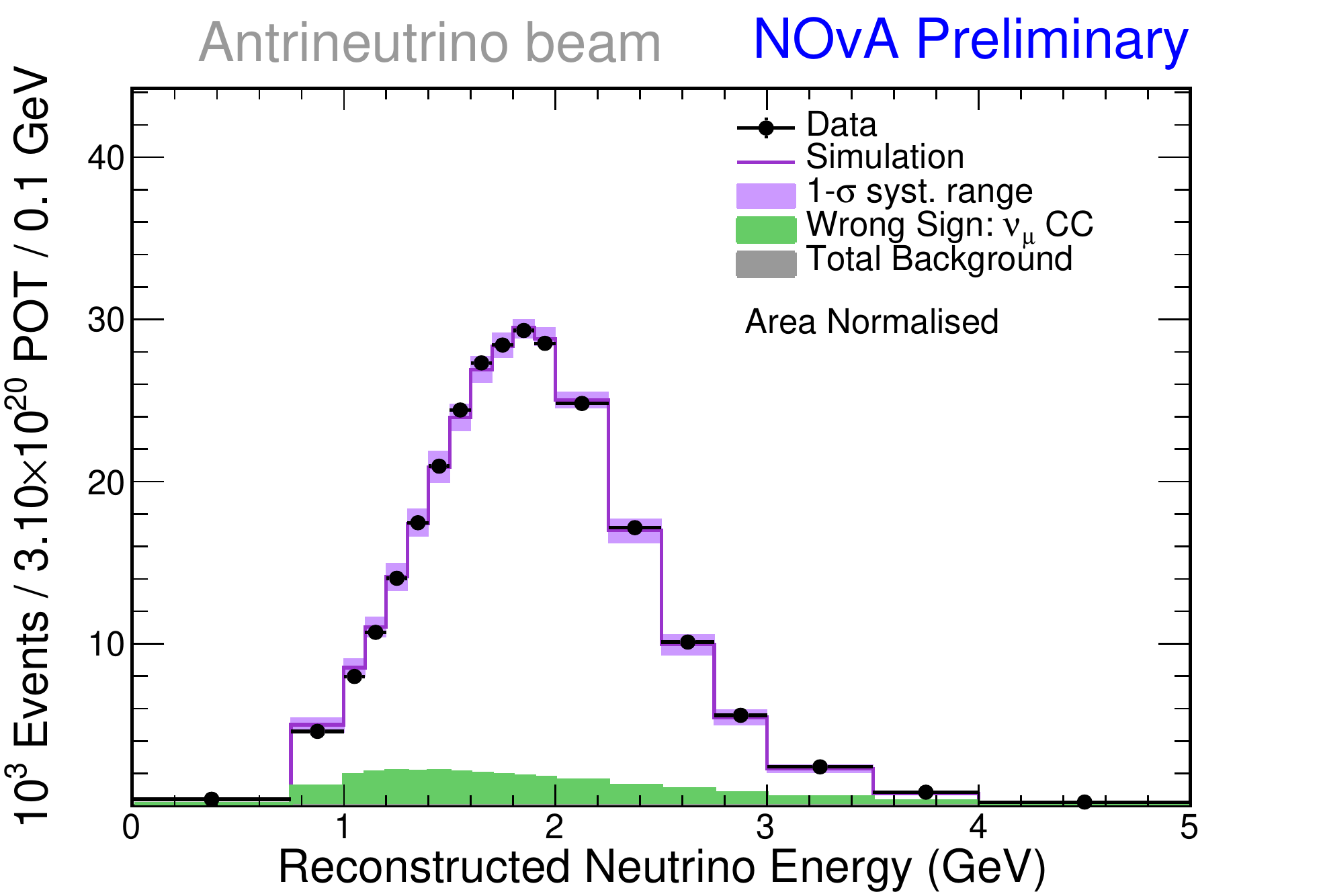}
    \caption{Reconstructed  $\nu_\mu$ (left) and $\bar{\nu}_\mu$ (right) energy spectra in ND Data and MC, merging all quartiles, area-normalized. Violet bands represent shape-only systematics uncertainties.}
    \label{fig:ndnumue}
    \end{figure}

The reconstructed energy distributions of the selected ND $\nu_\mu$ ($\bar{\nu}_\mu$) CC events in the neutrino (antineutrino) beam  data and MC  are shown in Figure~\ref{fig:ndnumue}.  All four $\nu_\mu$ energy quantiles are merged together to show the overall distribution. Data and MC are area-normalized. Violet bands represent shape-only systematic uncertainties. Data/MC differences in the POT normalization  are found to be 1.3\% and 0.5\% for $\nu_\mu$ and $\bar{\nu}_\mu$, respectively.  We normalize ND MC to Data in each energy quantile, then extrapolate the 4 quantiles to the FD and apply oscillations to predict the $\nu_\mu$ ($\bar{\nu}_\mu$ ) disappearance appearance signal spectra.

\begin{figure}[h]
    \centering
    \includegraphics[width=16cm]{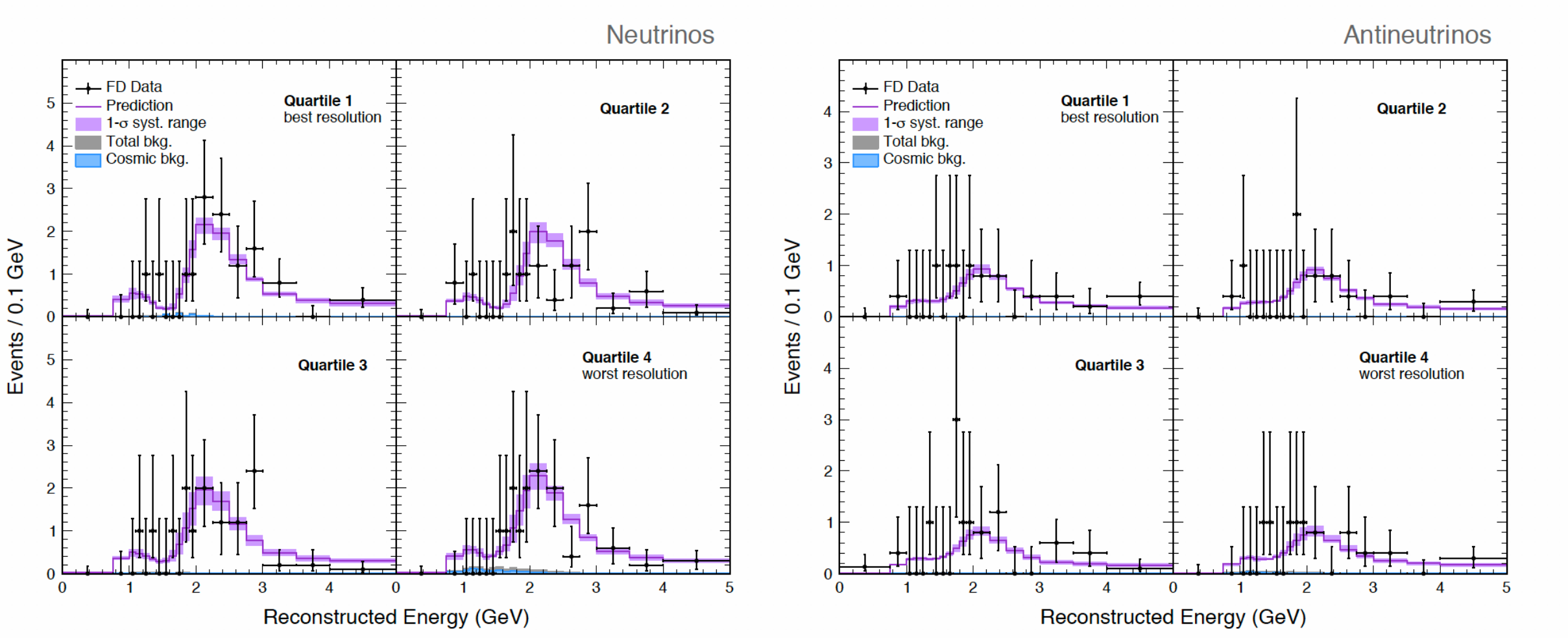}
    \caption{Reconstructed $\nu_\mu$ (left) and $\bar{\nu}_\mu$ (right) energy spectra in the four quartiles for the selected Data events and MC predictions in the FD. The MC predictions are extrapolated from the ND with the best fit oscillation parameters applied.}
    \label{fig:fdnumu4q}
    \end{figure}

The FD cosmic ray background is estimated from the FD cosmic data collected during the 550-$\mu$s NuMI spill trigger window, excluding the selection time window centered on the beam spill. The beam neutrino backgrounds for $\nu_\mu$ ($\bar{\nu}_\mu$) are estimated with FD simulation. Figure~\ref{fig:fdnumu4q} and \ref{fig:fdnumue} show the selected $\nu_\mu$ ($\bar{\nu}_\mu$) FD Data events and predictions in the individual four quartiles and in all quartiles.  The predictions in the four quartiles are extrapolated  from the ND independently, with the best $\nu_\mu$ ($\bar{\nu}_\mu$) and $\nu_e$ ($\bar{\nu}_e$) combined fit oscillation parameters applied. Adding up all energy quantiles, 113 (65) $\nu_\mu$ ($\bar{\nu}_\mu$) CC candidate events are observed from the FD neutrino (antineutrino) beam data.  We expect 731 (267) $\nu_\mu$ ($\bar{\nu}_\mu$)  events before oscillation in the FD and 125.2 (51.9) events with the best fit oscillation parameters applied. The predicted numbers of background events under the best fit parameters are 5.0 and 1.4 for neutrino and antineutrino beams, respectively.

  \begin{figure}[h]
    \centering
    \includegraphics[width=5.5cm]{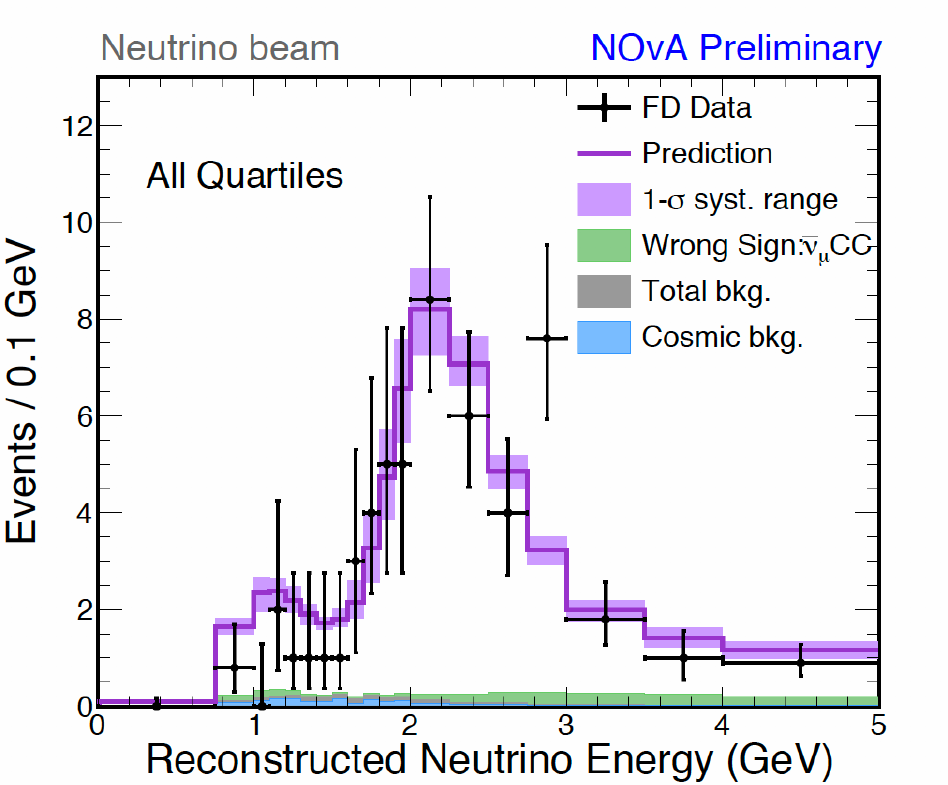}
     \includegraphics[width=5.5cm]{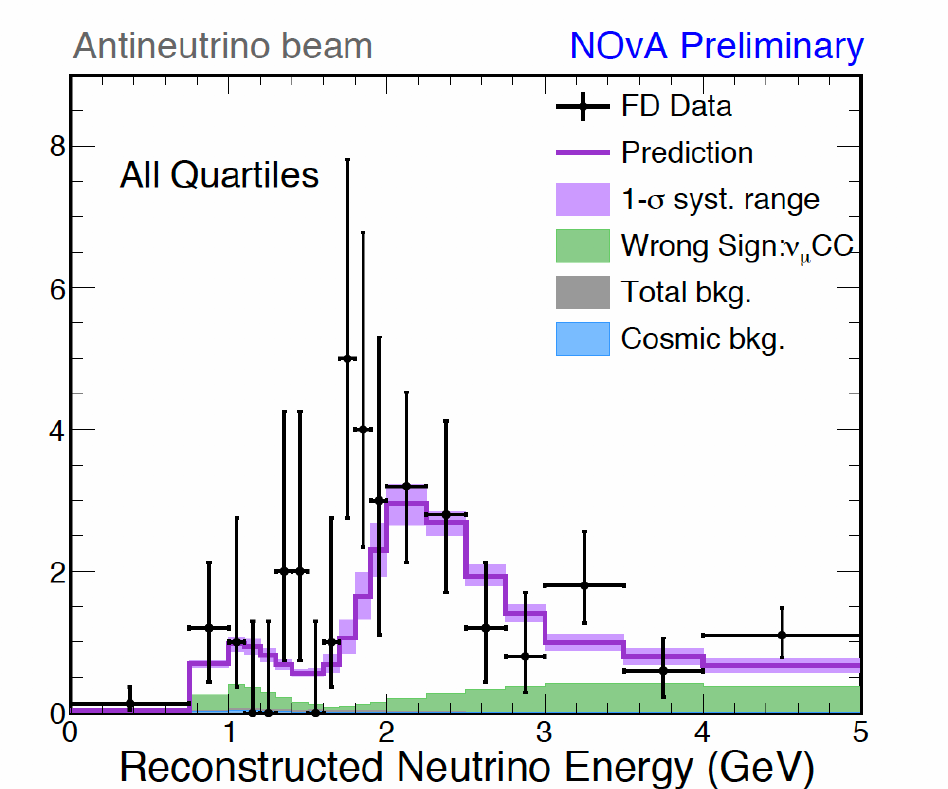}
    \caption{Reconstructed $\nu_\mu$ (left) and $\bar{\nu}_\mu$ (right) energy spectra in all quartiles for the selected data events and predictions in the FD. The predictions are extrapolated from the ND with the best fit oscillation parameters applied.}
    \label{fig:fdnumue}
    \end{figure}

\begin{figure}[h]
    \centering
    \includegraphics[width=8cm]{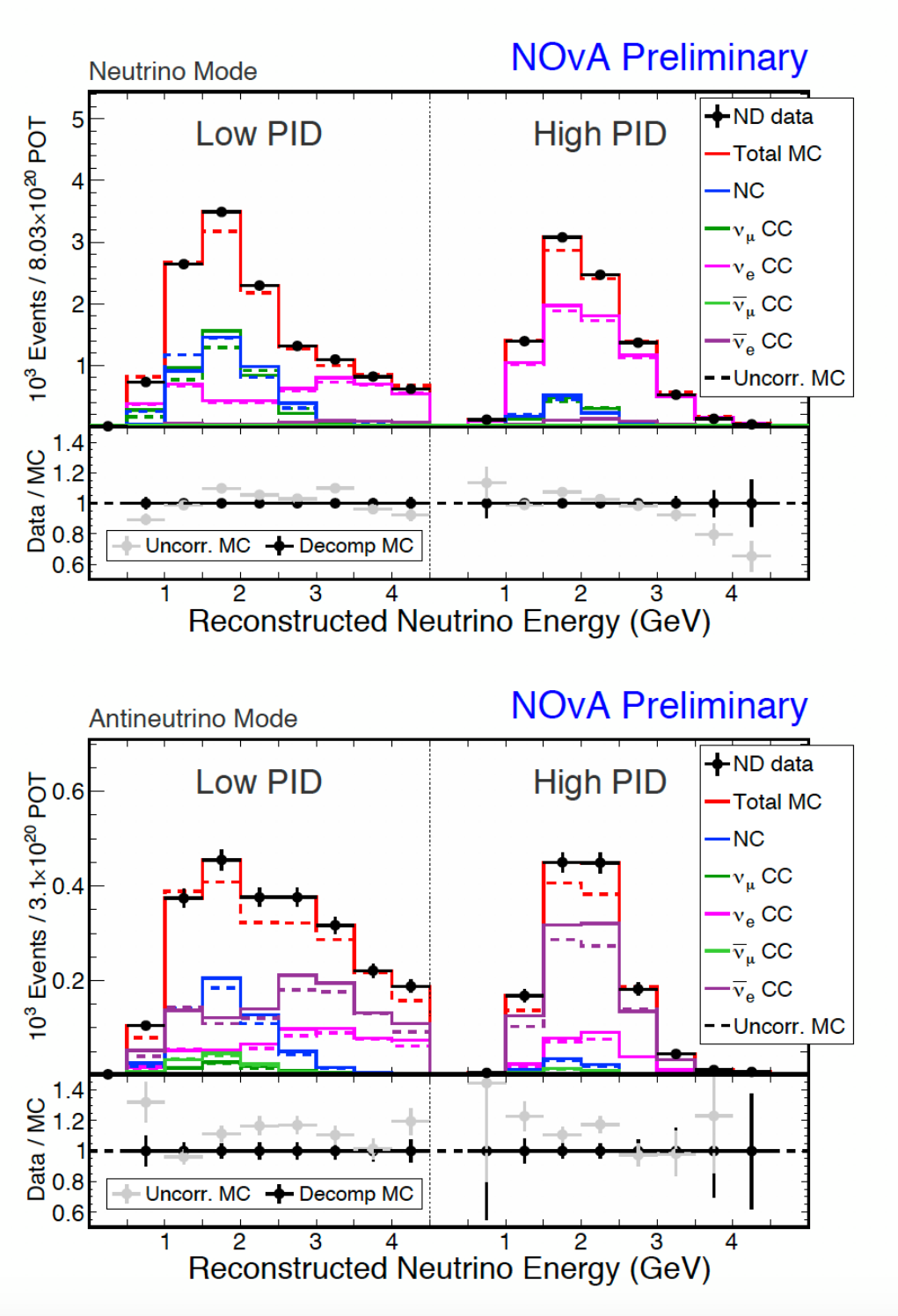}
    \caption{Reconstructed  $\nu_e$ (top) and $\bar{\nu}_e$ (bottom) energy spectra in ND data and MC. Dashed lines represent MC spectra before the beam and cross-section correction applied. Violet bands represent shape-only systematics uncertainties.}
    \label{fig:ndnuee}
    \end{figure}

The reconstructed energy distributions of the selected ND $\nu_e$ ($\bar{\nu}_e$) CC events in the neutrino (antineutrino) beam  Data and MC  are shown in Figure~\ref{fig:ndnuee}. We analyze the $\nu_e$ ($\bar{\nu}_e$) energy spectra in two PID bins ("low PID" and "high PID") to take advantage of the isolated high purity sample in the high PID bin. For the neutrino beam, contained and uncontained $\nu_\mu$ events are used to tune the pion and kaon contributions to the beam $\nu_e$, and Michel electrons are used to constrain NC/$\nu_\mu$-CC balance in each neutrino energy bin.  For the antineutrino beam, because of the low statistics, we scale all beam background components evenly to match the data. Each background component in the ND is propagated independently in energy and PID bins from the ND to the FD to predict backgrounds.

The FD cosmic ray background in the $\nu_e$ ($\bar{\nu}_e$) analysis is also estimated from the FD data in the sideband of the spill time window.  The $\nu_e$ ($\bar{\nu}_e$) appearance signal is predicted by applying oscillation parameters to the FD $\nu_e$ ($\bar{\nu_e}$) CC simulation with the energy taken from the unoscillated  FD $\nu_\mu$ ($\bar{\nu_\mu}$)  energy spectrum extrapolated from the ND. Figure~\ref{fig:fdnumue} shows the selected $\nu_e$ ($\bar{\nu}_e$) FD data events and FD predictions in the low PID bin, the high PID bin and the peripheral region. In total, 58 (18) $\nu_e$ ($\bar{\nu}_e$) CC candidate events are observed from the FD neutrino (antineutrino) beam data.  We expect 59.0 (15.9) $\nu_e$ ($\bar{\nu}_e$)  in the FD with the best fit oscillation parameters applied. The predicted background yields are 15.1 and 5.3 for the $\nu_e$ and $\bar{\nu}_e$ measurements, dominated by beam $\nu_e$ and $\bar{\nu}_e$. Wrong sign backgrounds depend on the oscillation parameters and are predicted to be less than one event for both $\nu_e$ and $\bar{\nu}_e$ appearances. Upon the observed 18 FD $\bar{\nu}_e$ candidates and the predicted of 5.3 background events, the observation corresponds to a 4.2 $\sigma$ excess over the background prediction, which is the first evidence of $\bar{\nu}_e$  appearance in long-baseline neutrino experiments.

  \begin{figure}[t]
    \centering
    \includegraphics[width=6.5cm]{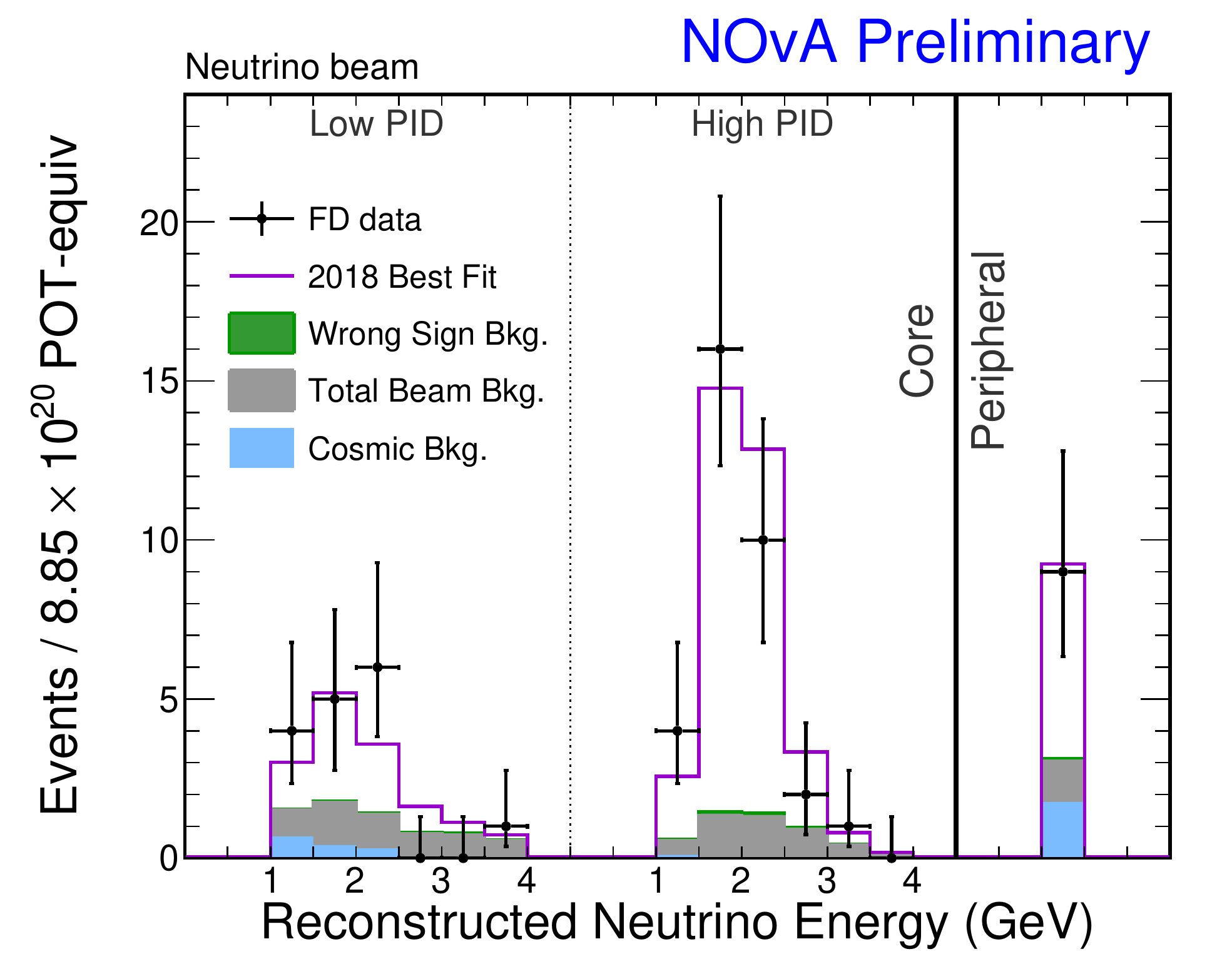}
     \includegraphics[width=6.5cm]{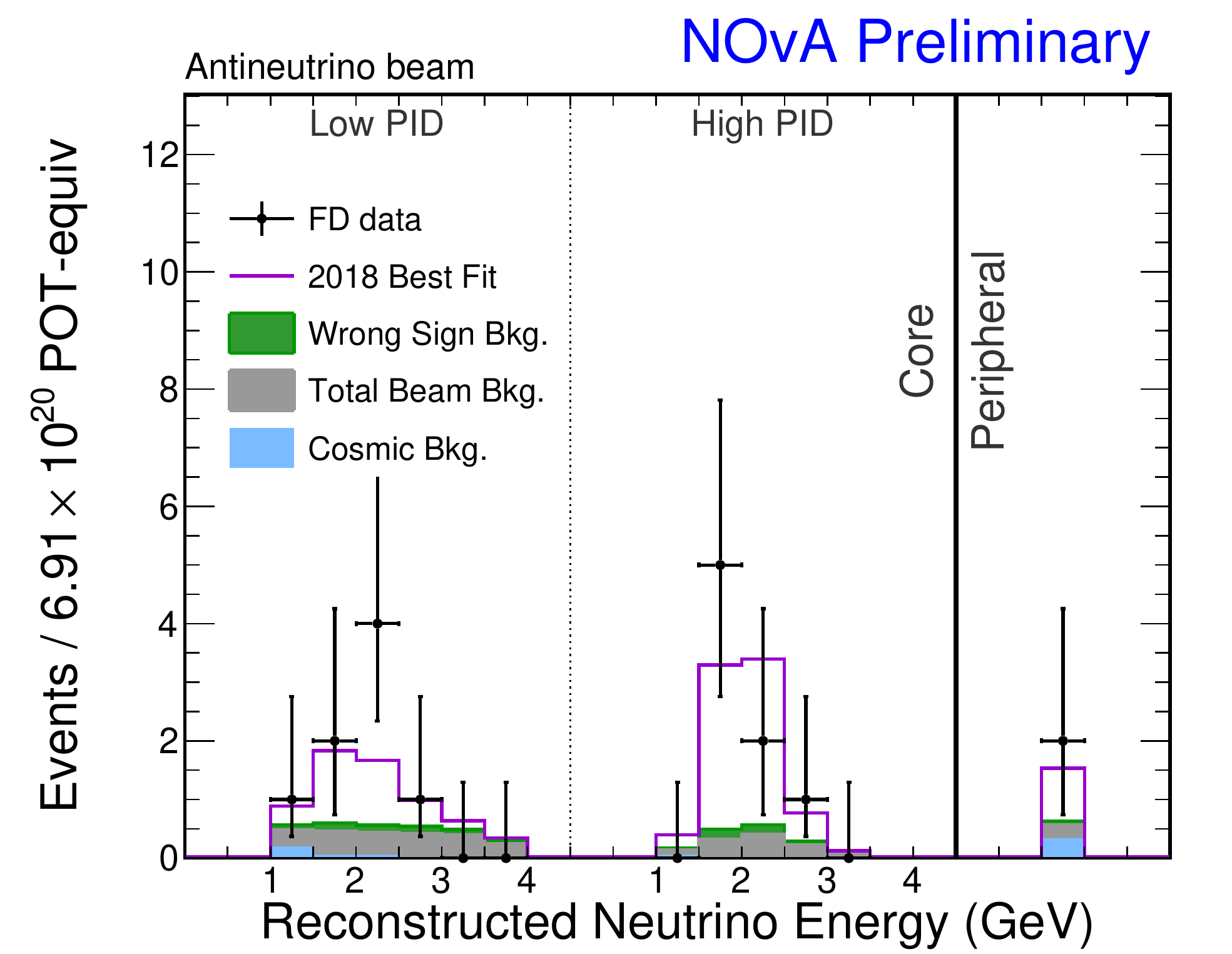}
    \caption{Reconstructed $\nu_e$ (left) and $\bar{\nu}_e$ (right) energy spectra in the low and high PID bins for the selected data events and predictions in the FD, along with event counts in the peripheral region. The predictions are extrapolated from the ND with the best fit oscillation parameters applied.}
    \label{fig:fdnumue}
    \end{figure}

\begin{figure}[h]
    \centering
    \includegraphics[width=6.5cm]{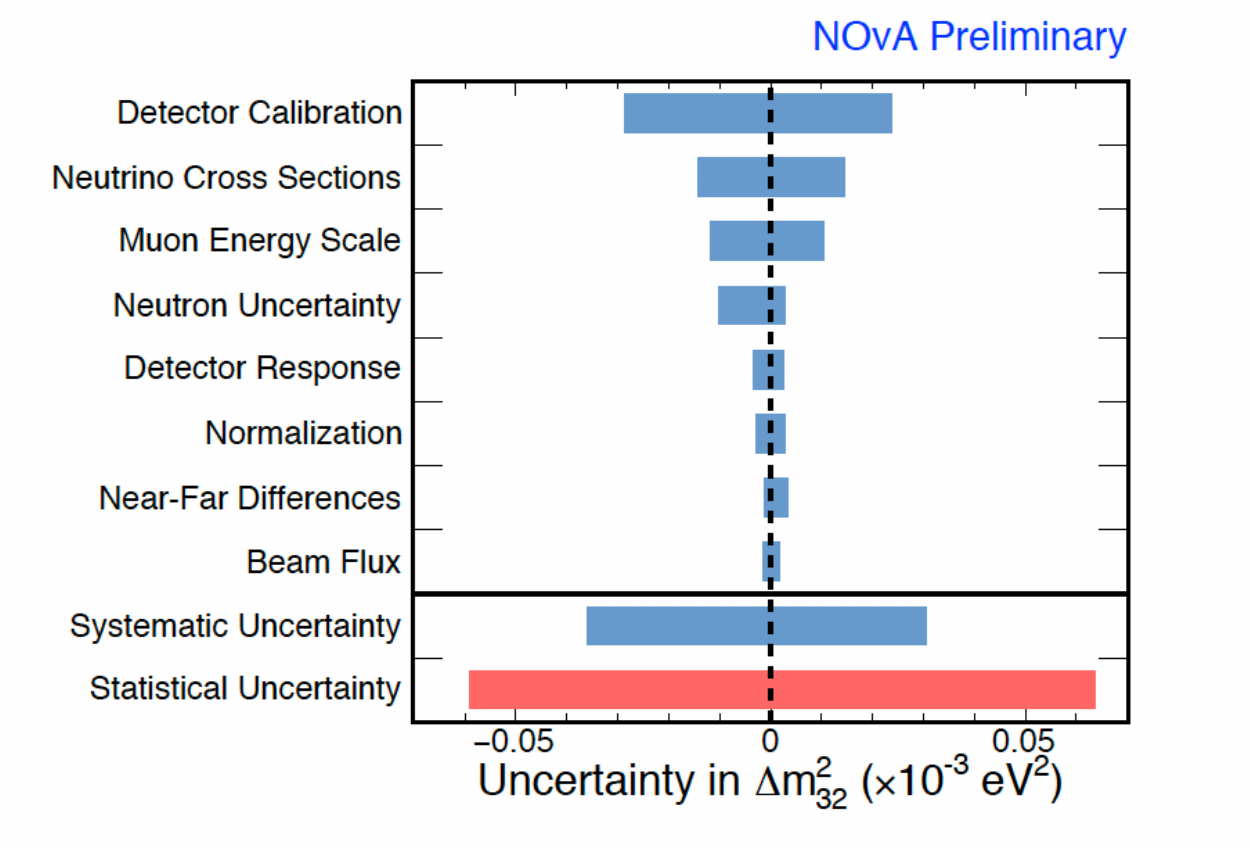}
     \includegraphics[width=6.5cm]{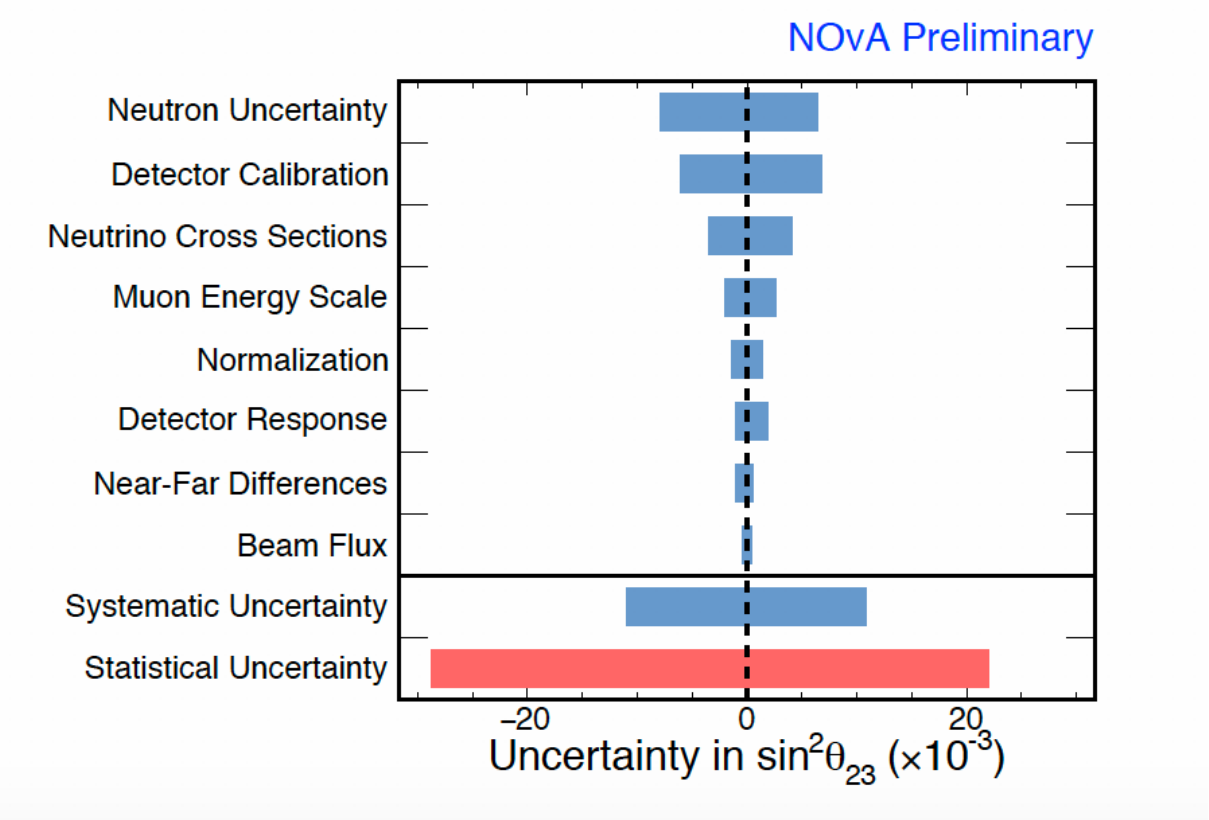}
     \includegraphics[width=6.5cm]{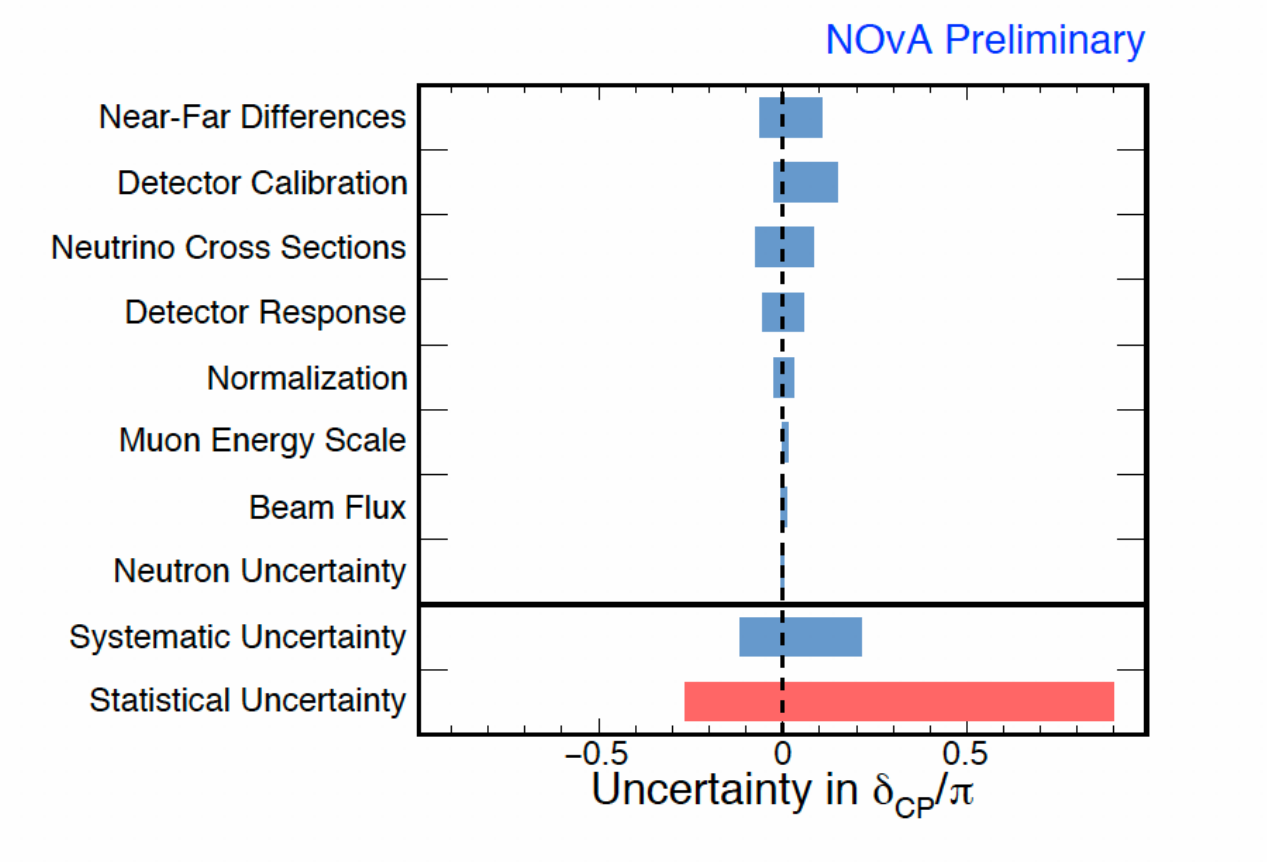}
    \caption{Systematic and statistical uncertainties for $\Delta m^2_{32}$, $\sin^2\theta_{23}$ and $\delta_{CP}$.}
    \label{fig:sys}
    \end{figure}

%\newpage
The extrapolation technique eliminates most of the systematic uncertainties. Remaining systematic uncertainties after the extrapolation are evaluated by extrapolating ND data with nominal MC and systematically modified MC samples, with variations to the normalization (POT counting), neutrino cross-sections, neutron simulation, beam flux simulation, detector calibration, neutrino energy scales, non-linearity in detector responses and other smaller uncertainties. Figure~\ref{fig:sys} shows systematic and statistical uncertainties for $\Delta m^2_{32}$, $\sin^2\theta_{23}$ and $\delta_{CP}$. Detector Calibration, neutrino cross sections, and muon energy scale are found to have large impacts on the results. Neutron simulation is a large source of systematic uncertainties in the $\bar{\nu}_\mu$ and $\bar{\nu}_e$ analyses. The upcoming NOvA test beam program will reduce the calibration and detector response uncertainties. 

\section{Joint Appearance and Disappearance Fit}

Energy distributions in the $\nu_\mu$ ( $\bar{\nu}_\mu$) quantiles and $\nu_e$ ($\nu_e$) PID bins are fit to data to extract oscillation parameters. In this combined fit, oscillation parameters $\sin^2\theta_{13}$ ($0.021\pm0.001$), $\sin^2\theta_{12}$ ($0.307^{+0.013}_{-0.012}$) and $\Delta m^2_{21} (7.53\pm 0.18 \times 10^{-5}$eV$^2$) are constrained according other experiments. The systemic uncertainties are included as nuisance parameters in the fit. 

\begin{figure}[h]
\centering
\includegraphics[width=7cm]{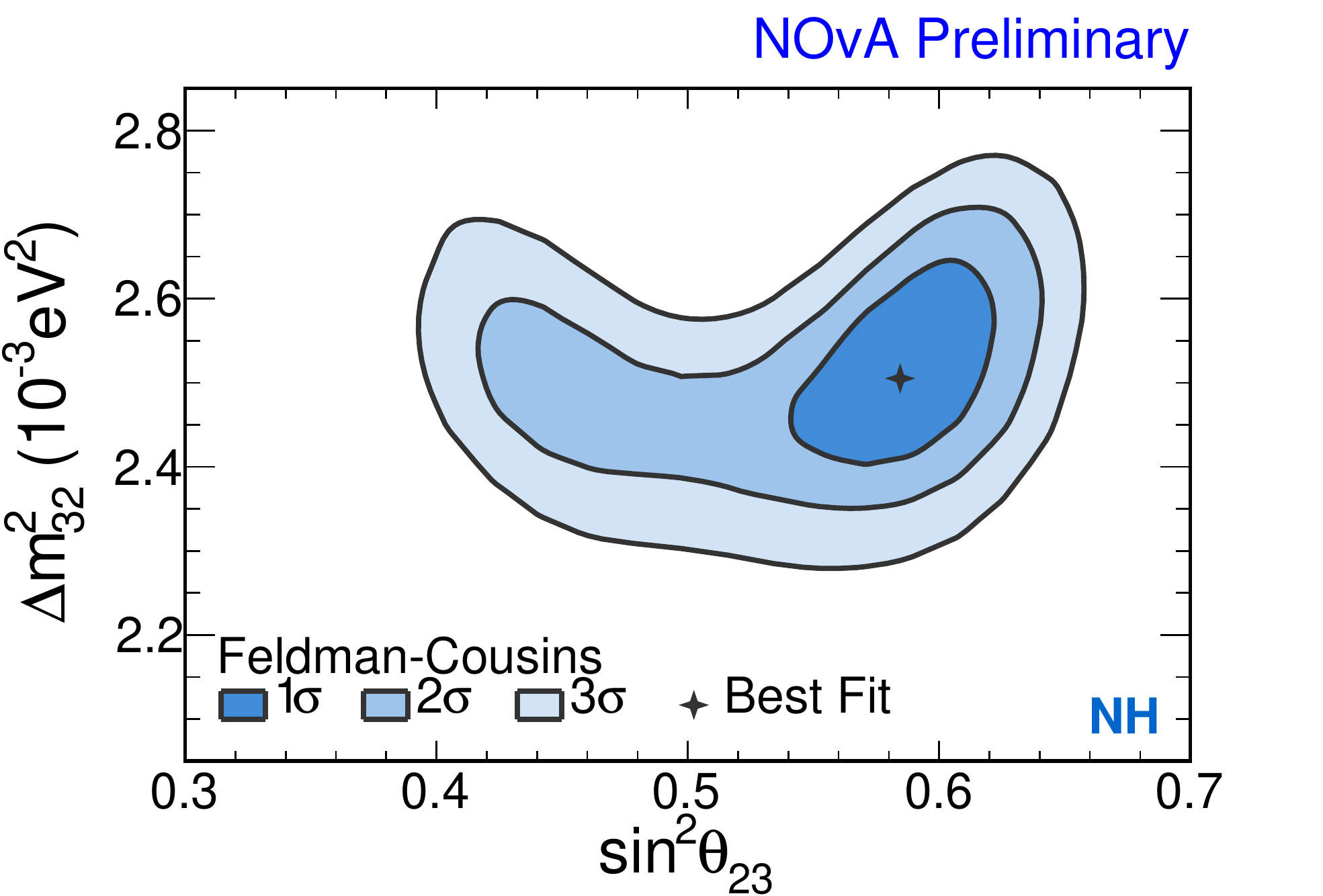}
\includegraphics[width=7cm]{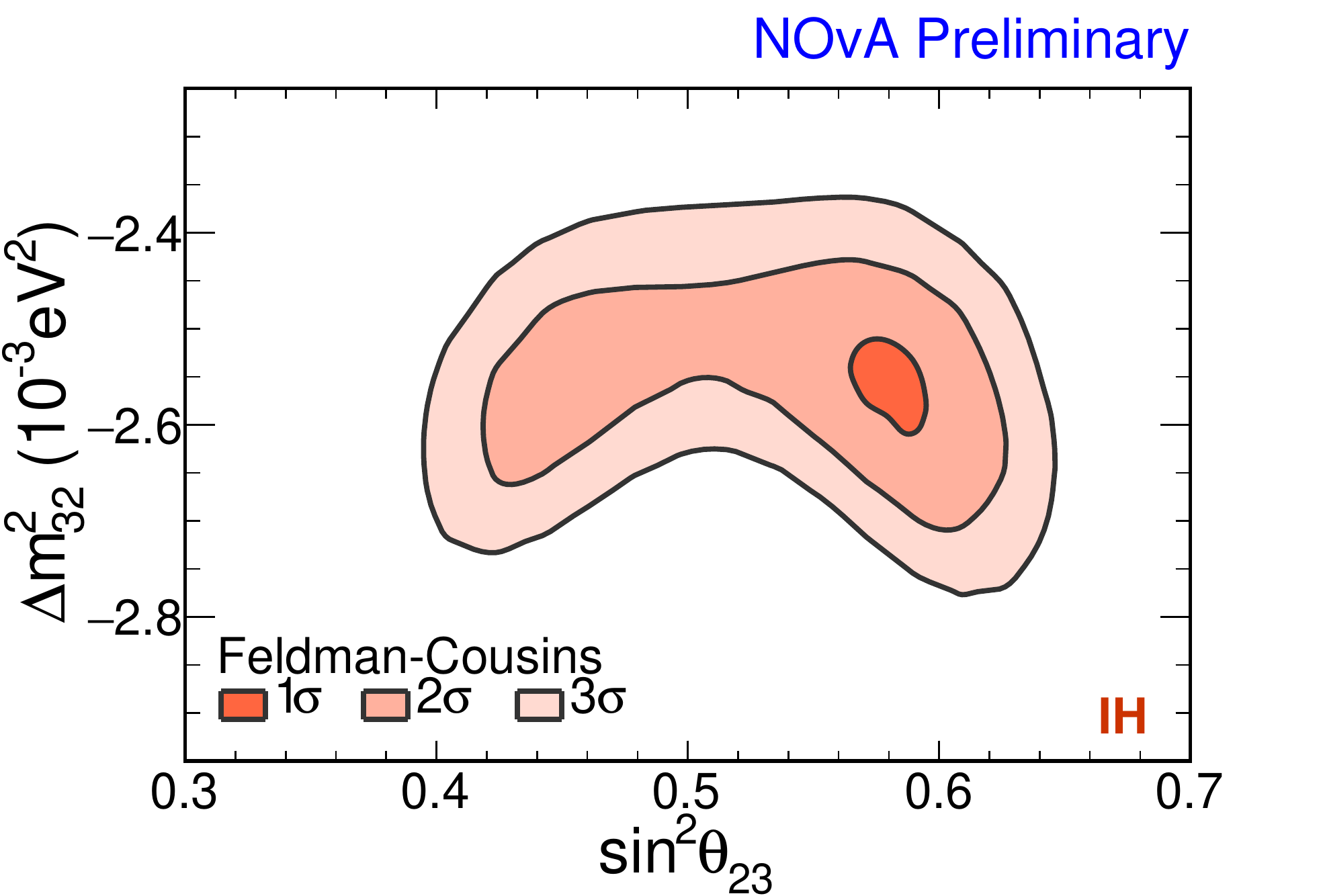}
\includegraphics[width=7cm]{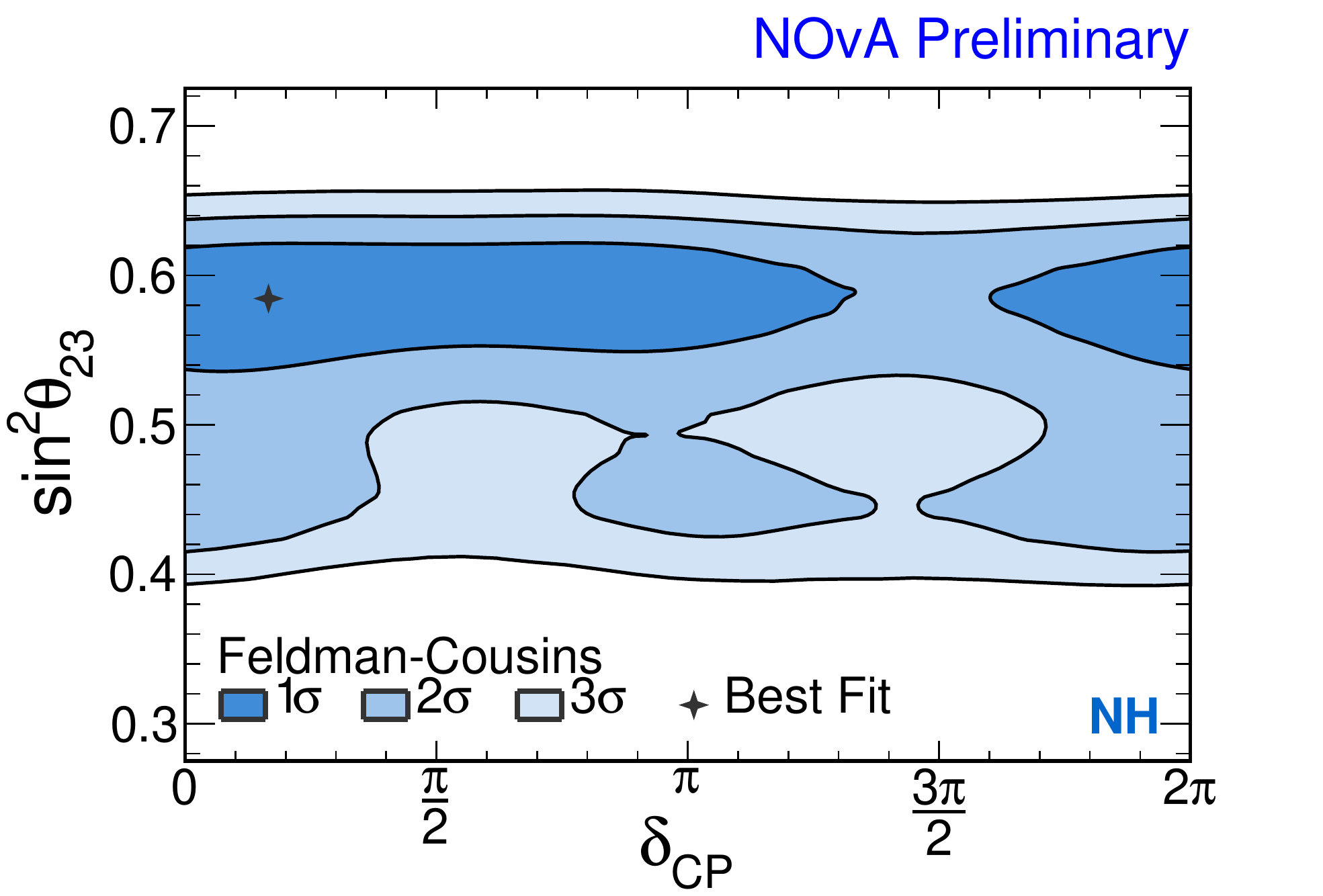}
\includegraphics[width=7cm]{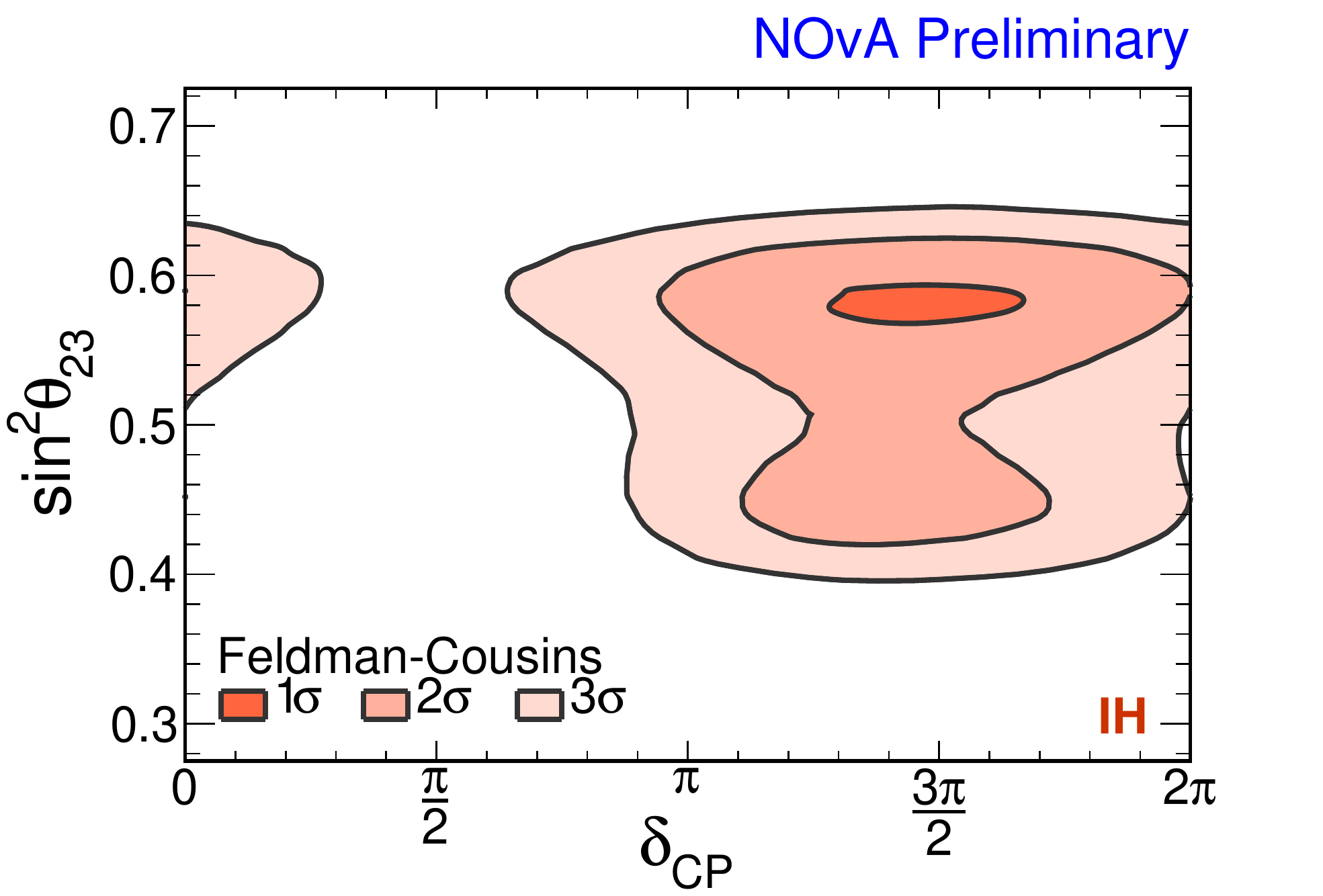}
\caption{Allowed regions of $\Delta m^2_{32}$ vs. $\sin^2\theta_{23}$ (top) and $\sin^2\theta_{23}$ vs. $\delta_{CP}$ (bottom) at 1, 2, and $3\sigma$ for each of the hierarchies, produced by the joint fit.}\label{fig:contour}
\end{figure}

\begin{figure}[h]
\centering
\includegraphics[width=7cm]{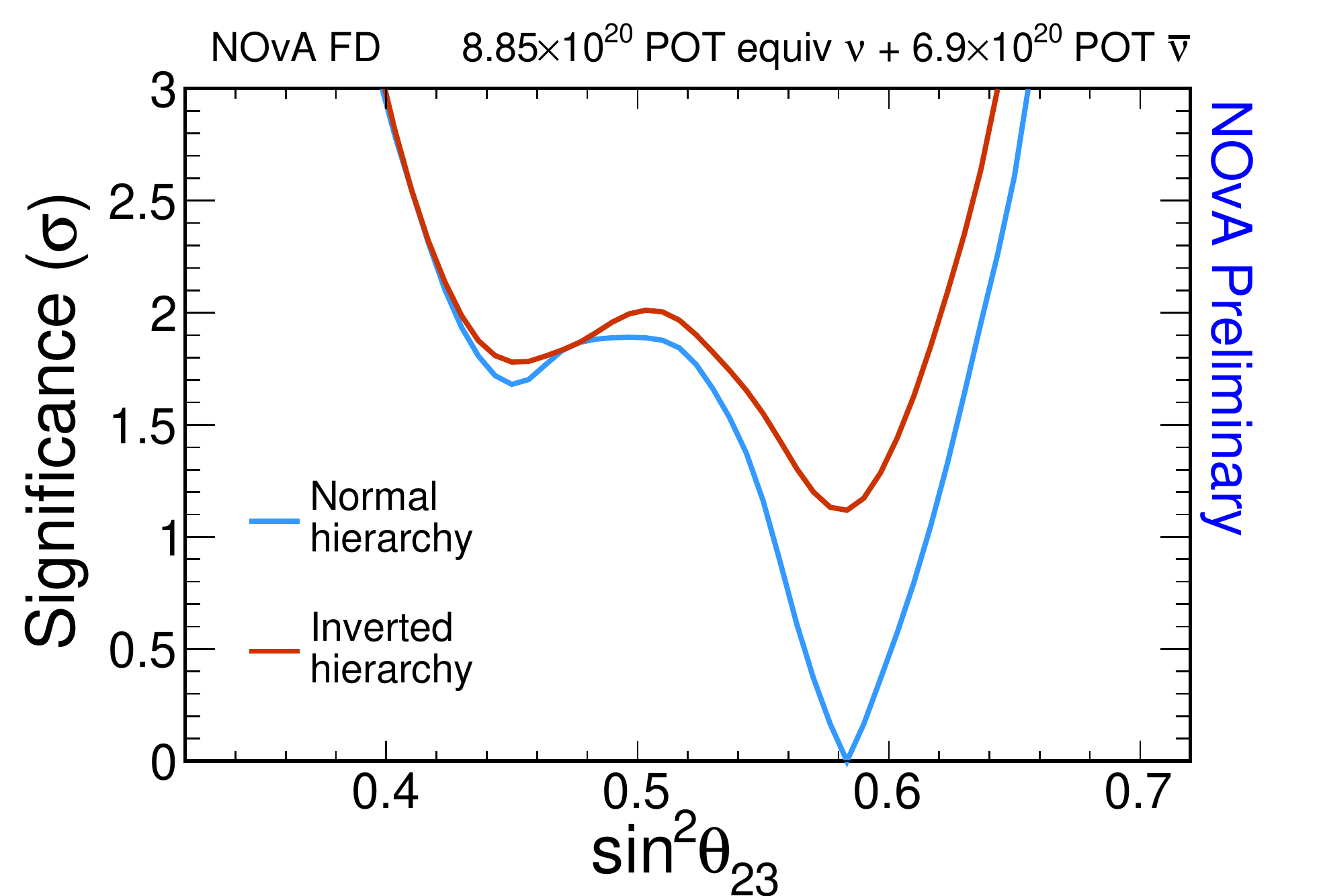}
\includegraphics[width=7cm]{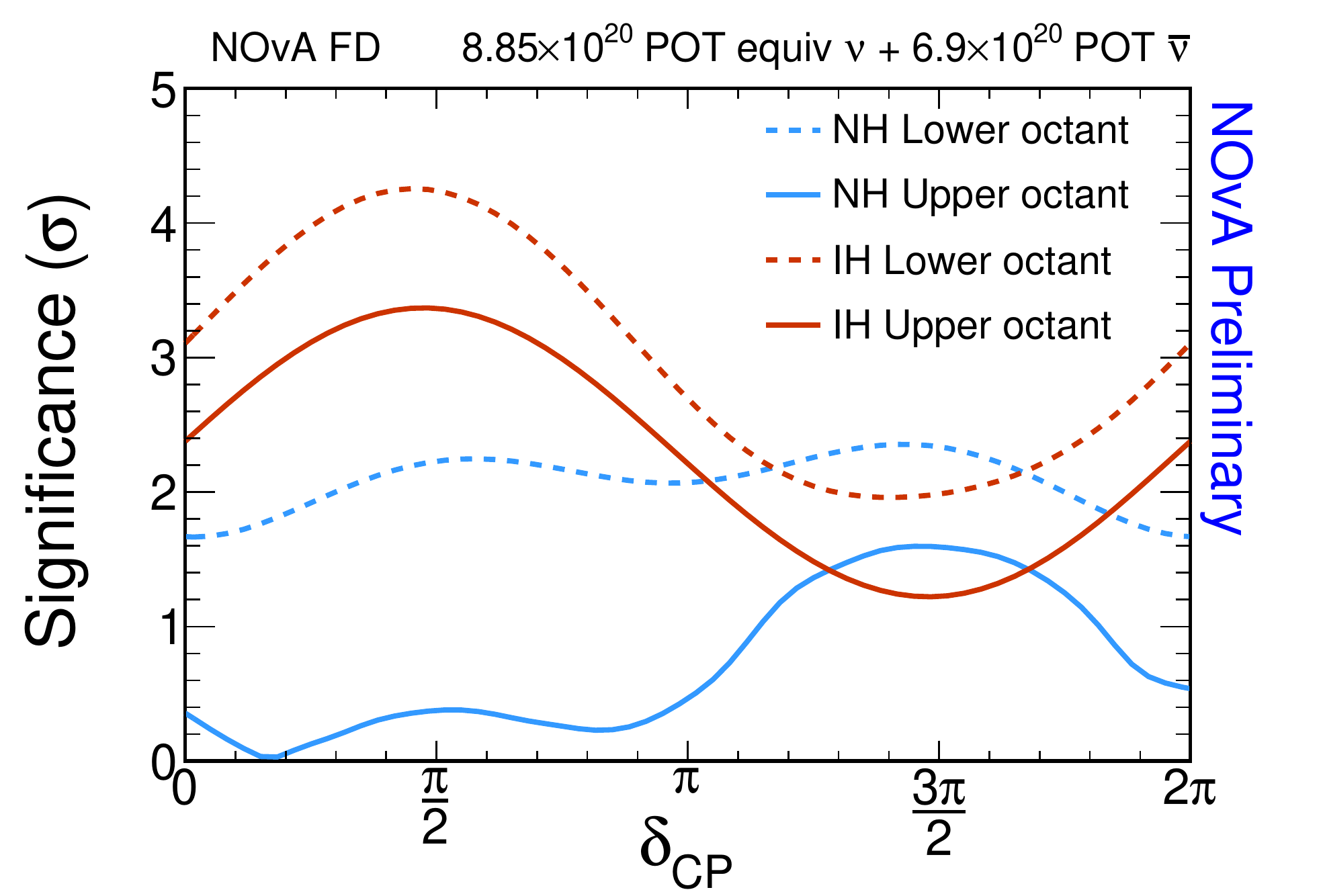}
\caption{(left) Significances as functions of $\Delta m^2_{32}$,  profiling over  $\sin^22\theta_{13}, \sin^2\theta_{23}, \delta_{CP}$  and systematics, (right) Significances as functions of $\delta_{CP}$, profiling over  $\sin^22\theta_{13}, \Delta m^2_{32}, \sin^2\theta_{23}$  and systematics. Feldman-Cousins corrections are applied.}\label{fig:fcslices}
\end{figure}

The resulting allowed regions of $\Delta m^2_{32}$ vs. $\sin^2\theta_{23}$ and $\sin^2\theta_{23}$ vs. $\delta_{CP}$ at 1, 2, and $3\sigma$ for each of the hierarchies produced by the joint fit are shown in Figure~\ref{fig:contour}. Feldman-Cousins (FC) corrections are used to obtain these contours to interpret the measured $-2\Delta \ln(L)$ in the data~\cite{Feldman:1997qc}. The 2-D FC contours are converted into significances as functions of $\sin^2\theta_{23}$ and $\delta_{CP}$ by profiling over oscillation parameters and systematics, as shown in Figure~\ref{fig:fcslices}.

The global best fit gives us normal hierarchy, $\delta_{CP}=0.17\pi$, $\sin^2\theta_{23} = 0.58\pm0.03$ (upper octant) and  $\Delta m^2_{32}=(2.51^{+0.12}_{-0.08})\times10^{-3} \rm{eV}^2$. The results prefer the normal hierarchy at 1.8$\sigma$ and consistent with all $\delta_{CP}$ values in the normal hierarchy at within $1.6\sigma$. The data also prefer a non-maximal mixing at $1.8\sigma$, favoring the upper octant at similar level. The combined fit excludes values around $\delta_{CP}=\pi/2$ in the inverted mass hierarchy by more than $3\sigma$.

\section{Summary and Future Plan}

\begin{figure}[h]
\centering
\includegraphics[width=7cm]{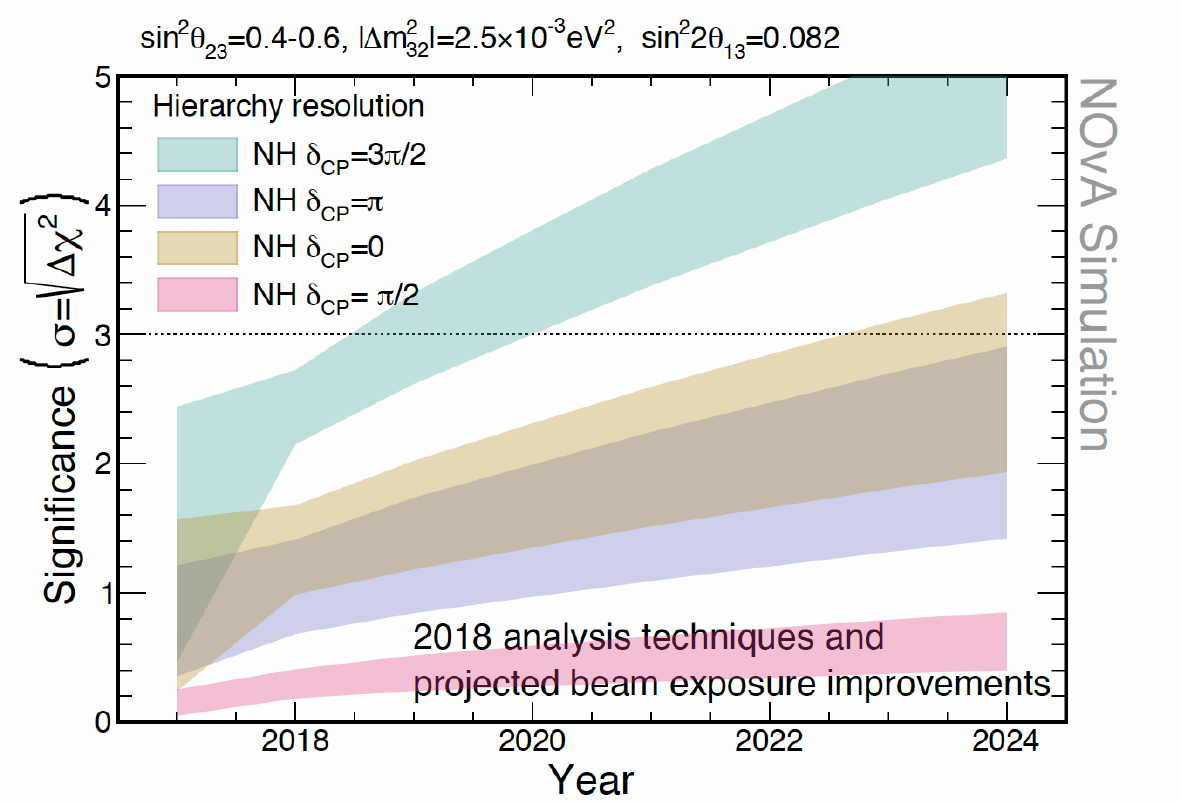}
\includegraphics[width=7cm]{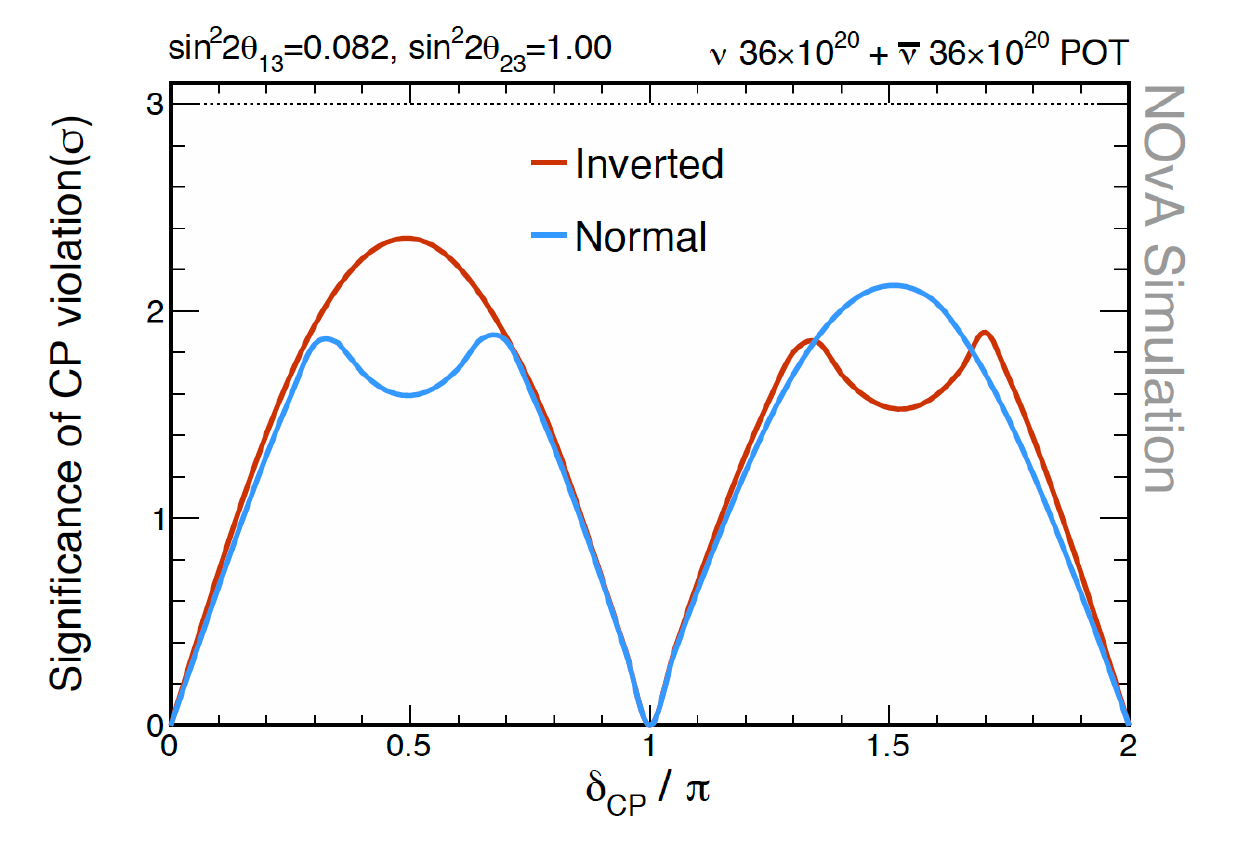}
\caption{(left) Expected significances of the mass hierarchy at NOvA from 2017 to 2024 under different $\delta_{CP}$ hypotheses. (right) Expected significances as a function of $\delta_{CP}$ with the full NOvA data.}\label{fig:novareach}
\end{figure}

With $8.85 \times10^{20}$ ($6.9 \times10^{20}$) of POT NuMI neutrino (anti-neutrino) beam data, we have performed NOvA's first FHC+RHC combined oscillation analysis. The significance of the $\bar{\nu}_e$ appearance is greater than $4\sigma$, and our data prefers the normal mass hierarchy and the upper octant. NOvA has been taking antineutrino data since 2017, and the NuMI beam will be switched back to generate neutrinos in 2019. After that NuMI will run for $50\%$ neutrino data and $50\%$ anti-neutrino data. NOvA will extend running through 2024 to increase statistics, and the ongoing test beam program and proposed accelerator improvements could enhance the ultimate reach. The expected significances of the mass hierarchy and CP violation at NOvA are shown in Figure~\ref{fig:novareach}. If the value of $\delta_{CP}$ is close to $3\pi/2$, as hinted by T2K~\cite{Abe:2015awa}, the significance of the mass hierarchy at NOvA will reach $3\sigma$ by 2020 and $5\sigma$ by 2024.

\end{document}